\documentclass[aps,prb,twocolumn,superscriptaddress,amsmath,amssymb]{revtex4}
\usepackage{graphicx}
\usepackage{hyperref}

\begin{document}

\title{Nonlinear uniaxial pressure dependence of $T_c$ in iron-based superconductors}
\author{Zhaoyu Liu}
\affiliation{Beijing National Laboratory for Condensed Matter Physics, Institute of Physics, Chinese Academy of Sciences, Beijing 100190, China}
\affiliation{School of Physical Sciences, University of Chinese Academy of Sciences, Beijing 100190, China}
\affiliation{Department of Physics, University of Washington, Seattle, WA 98195, USA}
\author{Yanhong Gu}
\affiliation{Beijing National Laboratory for Condensed Matter Physics, Institute of Physics, Chinese Academy of Sciences, Beijing 100190, China}
\affiliation{School of Physical Sciences, University of Chinese Academy of Sciences, Beijing 100190, China}
\author{Wenshan Hong}
\affiliation{Beijing National Laboratory for Condensed Matter Physics, Institute of Physics, Chinese Academy of Sciences, Beijing 100190, China}
\affiliation{School of Physical Sciences, University of Chinese Academy of Sciences, Beijing 100190, China}
\author{Tao Xie}
\affiliation{Beijing National Laboratory for Condensed Matter Physics, Institute of Physics, Chinese Academy of Sciences, Beijing 100190, China}
\affiliation{School of Physical Sciences, University of Chinese Academy of Sciences, Beijing 100190, China}
\author{Dongliang Gong}
\affiliation{Beijing National Laboratory for Condensed Matter Physics, Institute of Physics, Chinese Academy of Sciences, Beijing 100190, China}
\affiliation{School of Physical Sciences, University of Chinese Academy of Sciences, Beijing 100190, China}
\author{Xiaoyan Ma}
\affiliation{Beijing National Laboratory for Condensed Matter Physics, Institute of Physics, Chinese Academy of Sciences, Beijing 100190, China}
\affiliation{School of Physical Sciences, University of Chinese Academy of Sciences, Beijing 100190, China}
\author{Jing Liu}
\affiliation{Beijing National Laboratory for Condensed Matter Physics, Institute of Physics, Chinese Academy of Sciences, Beijing 100190, China}
\affiliation{School of Physical Sciences, University of Chinese Academy of Sciences, Beijing 100190, China}
\author{Cheng Hu}
\affiliation{Beijing National Laboratory for Condensed Matter Physics, Institute of Physics, Chinese Academy of Sciences, Beijing 100190, China}
\affiliation{School of Physical Sciences, University of Chinese Academy of Sciences, Beijing 100190, China}
\author{Lin Zhao}
\affiliation{Beijing National Laboratory for Condensed Matter Physics, Institute of Physics, Chinese Academy of Sciences, Beijing 100190, China}
\affiliation{School of Physical Sciences, University of Chinese Academy of Sciences, Beijing 100190, China}
\author{Xingjiang Zhou}
\affiliation{Beijing National Laboratory for Condensed Matter Physics, Institute of Physics, Chinese Academy of Sciences, Beijing 100190, China}
\affiliation{School of Physical Sciences, University of Chinese Academy of Sciences, Beijing 100190, China}
\affiliation{Songshan Lake Materials Laboratory , Dongguan, Guangdong 523808, China}
\author{R. M. Fernandes}
\affiliation{School of Physics and Astronomy, University of Minnesota, Minnesota, Minneapolis 55455, USA}
\author{Yi-feng Yang}
\affiliation{Beijing National Laboratory for Condensed Matter Physics, Institute of Physics, Chinese Academy of Sciences, Beijing 100190, China}
\affiliation{School of Physical Sciences, University of Chinese Academy of Sciences, Beijing 100190, China}
\affiliation{Songshan Lake Materials Laboratory , Dongguan, Guangdong 523808, China}
\author{Huiqian Luo}
\affiliation{Beijing National Laboratory for Condensed Matter Physics, Institute of Physics, Chinese Academy of Sciences, Beijing 100190, China}
\affiliation{Songshan Lake Materials Laboratory , Dongguan, Guangdong 523808, China}
\author{Shiliang Li}
\email{slli@iphy.ac.cn}
\affiliation{Beijing National Laboratory for Condensed Matter Physics, Institute of Physics, Chinese Academy of Sciences, Beijing 100190, China}
\affiliation{School of Physical Sciences, University of Chinese Academy of Sciences, Beijing 100190, China}
\affiliation{Songshan Lake Materials Laboratory , Dongguan, Guangdong 523808, China}

\begin{abstract}
We have systematically studied the effects of in-plane uniaxial pressure $p$ on the superconducting transition temperature $T_c$ in many iron-based superconductors. The change of $T_c$ with $p$ is composed of linear and nonlinear components. The latter can be described as a quadratic term plus a much smaller fourth-order term. In contrast to the linear component, the nonlinear $p$ dependence of $T_c$ displays a pronounced in-plane anisotropy, which is similar to the anisotropic response of the resistivity to $p$. As a result, it can be attributed to the coupling between the superconducting and nematic orders, in accordance with the expectations of a phenomenological Landau theory. Our results provide direct evidences for the interplay between nematic fluctuations and superconductivity, which may be a common behavior in iron-based superconductors.
\end{abstract}


%\maketitle must follow title, authors, abstract, \pac
\maketitle
\section{introduction}
Nematicity has been found in both cuprates and iron-based superconductors, consisting of an electronic state that breaks the in-plane $C_4$ rotational symmetry of the underlying lattice \cite{Fradkin_review}. In the former, nematic order seems associated with the pseudogap state \cite{DaouR10,SatoY17}, where many other types of orders are also found, such as stripes and charge density waves \cite{KeimerB15}. The scenario is much simpler in iron-based superconductors, where nematicity typically appears together with antiferromagnetism and superconductivity in the phase diagram \cite{FernandesRM14}. In several materials, a putative nematic quantum critical point (QCP) has been proposed to exist around the optimal doping level \cite{KuoHH16,HosoiS16,ThorsmolleVK16,LiuZ16,GuY17}, suggesting a close interplay between nematicity and superconductivity. Interestingly, two-fold anisotropy in the magnetoresistivity has been reported in the vicinity of the superconducting transition of slightly overdoped Ba$_{1-x}$K$_x$Fe$_2$As$_2$, which suggests the formation of a nematic superconductor \cite{LiJ17}. Theoretically, it has been proposed that nematic fluctuations can induce attractive pairing interaction that may enhance or even lead to superconductivity \cite{Yamase13,LedererS15,MetlitskiMA15,SchattnerY16,PhilippTD16,LedererS17,Kang16}. However, direct experimental evidence for the interplay between superconductivity and nematic fluctuations in iron-based superconductors is scarce. 

To shed light on this issue, here we study how the superconducting transition temperature $T_c$ changes with the in-plane uniaxial pressure $p$.  It has already been shown that the values of $dT_c/dp$ for $p$ applied within the $ab$ plane is very different compared to $p$ applied along the c axis \cite{MeingastC12,BohmerAE12,BohmerAE13,BohmerAE15,WangL17}. This type of anisotropic behavior is expected as observed in many other superconductors with layered structures \cite{JinDS92,MaesatoM01,DixOM09,SteppkeA17}. Since the nematic order breaks the in-plane rotational symmetry, it is possible to observe in-plane anisotropic superconducting properties if the coupling between the nematic order and superconductivity is significant. Even in optimally and overdoped regimes where neither the antiferromagnetic (AF) order nor the nematic order exists, nematic fluctuations can still be strong, giving rise to strongly anisotropic responses in the presence of uniaxial strain $p$. This is indeed observed in the resistivity of the normal state \cite{LiuZ16}, and thus may also affect the uniaxial pressure dependence of $T_c$. One of the difficulties to single out the nematic contribution to the observed behavior of $T_c(p)$ arises from the fact that $p$ induces not only the shear lattice distortion that couples to nematicity, but also lattice distortions associated with other symmetries that do not couple linearly to the nematic order parameter \cite{Fisher17}. To disentangle these contributions, our strategy here is to compare $T_c(p)$ for pressures applied along the Fe-Fe direction, and along the Fe-As-Fe or Fe-Se-Fe direction. This is because the former is along the nematic direction and should exhibit more significant effects than the latter.

Following this idea, we systematically studied the uniaxial pressure dependence of $T_c$ in many iron-based superconductors for $p$ applied within the $ab$ plane. It is found that $T_c$ can be described as a fourth-order polynomial function of $p$. While the linear term is essentially unaffected by the direction of $p$, the second-order nonlinear term varies significantly along different directions and samples. Comparing the results with a phenomenological Landau model that includes the biquadratic coupling between the nematic and superconducting order parameters, we conclude that this quadratic term is associated with nematic fluctuations. Since these effects can also be found in heavily overdoped samples, our results provide key insights for the impact of nematic fluctuations on superconductivity in iron-based superconductors.

\section{experiments}

\begin{figure}
\includegraphics[width=\columnwidth]{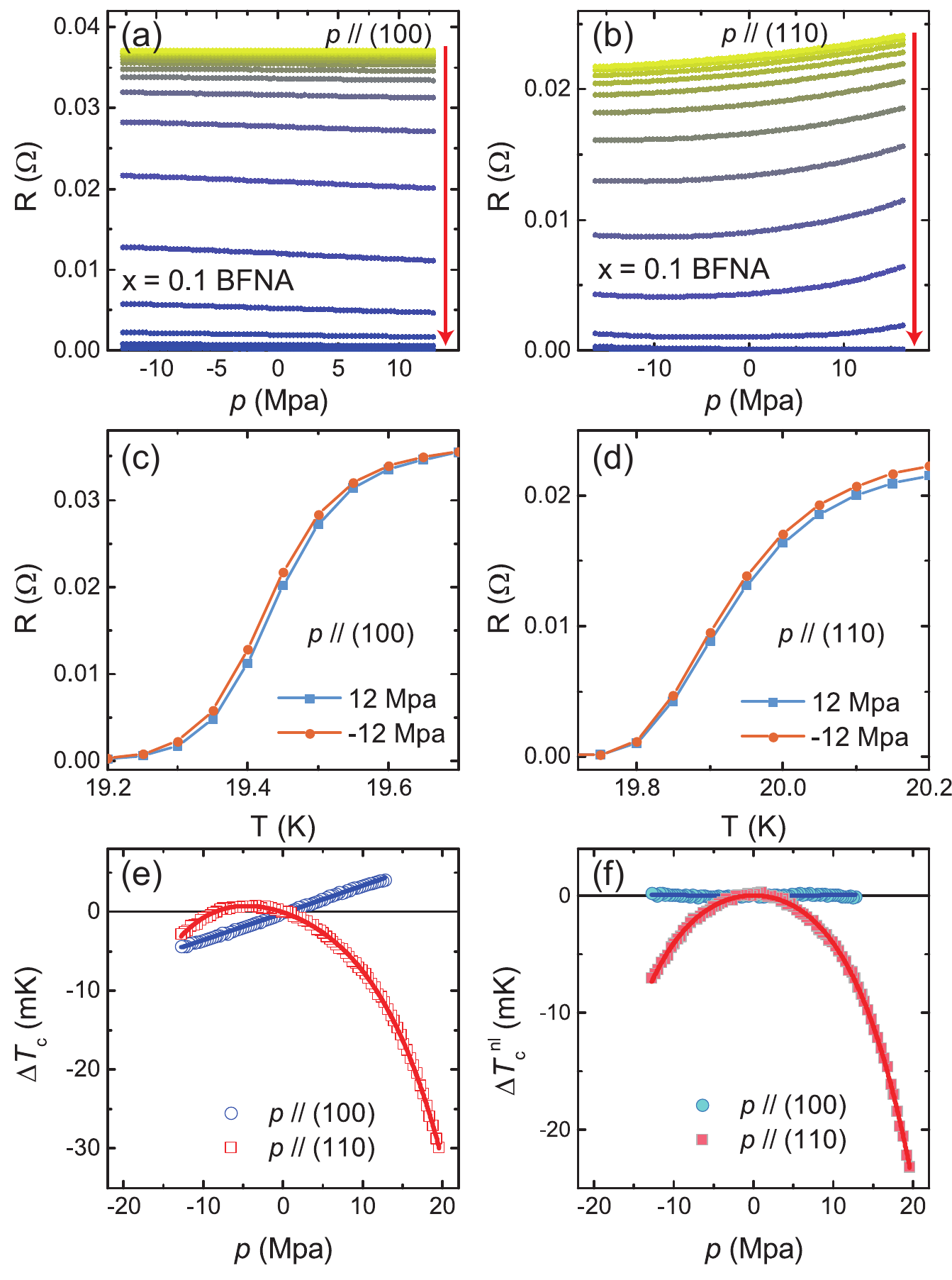}
\caption{(a) Uniaxial pressure dependence of the resistance for BaFe$_{1.9}$Ni$_{0.1}$As$_2$ with $p$ along the (100) direction. The temperature varies from 20 to 19.2 K with a step of 0.05 K as shown by the arrow. (b) Similar to (a) but for $p$ along the (110) direction. The temperature varies from 20.3 to 19.6 K with a step of 0.05 K. (c) \& (d) Converted temperature dependence of the resistance for $p$ along the (100) and (110) directions at $12$ and $-12$ MPa, respectively. The reason why $T_c$ of these two samples are different is because the current densities are different in the measurements due to their different cross sections. (e) Uniaxial pressure dependence of $\Delta T_c$ along the (100) (circles) and (110) (squares) directions. (f) Uniaxial pressure dependence of $\Delta T^{nl}_c$ along the (100) (circles) and (110) (squares) directions. The sold line for the case $p \parallel (110)$ is fitted as described in the main text.}
\label{fig1}
\end{figure}

\begin{figure}
\includegraphics[width=\columnwidth]{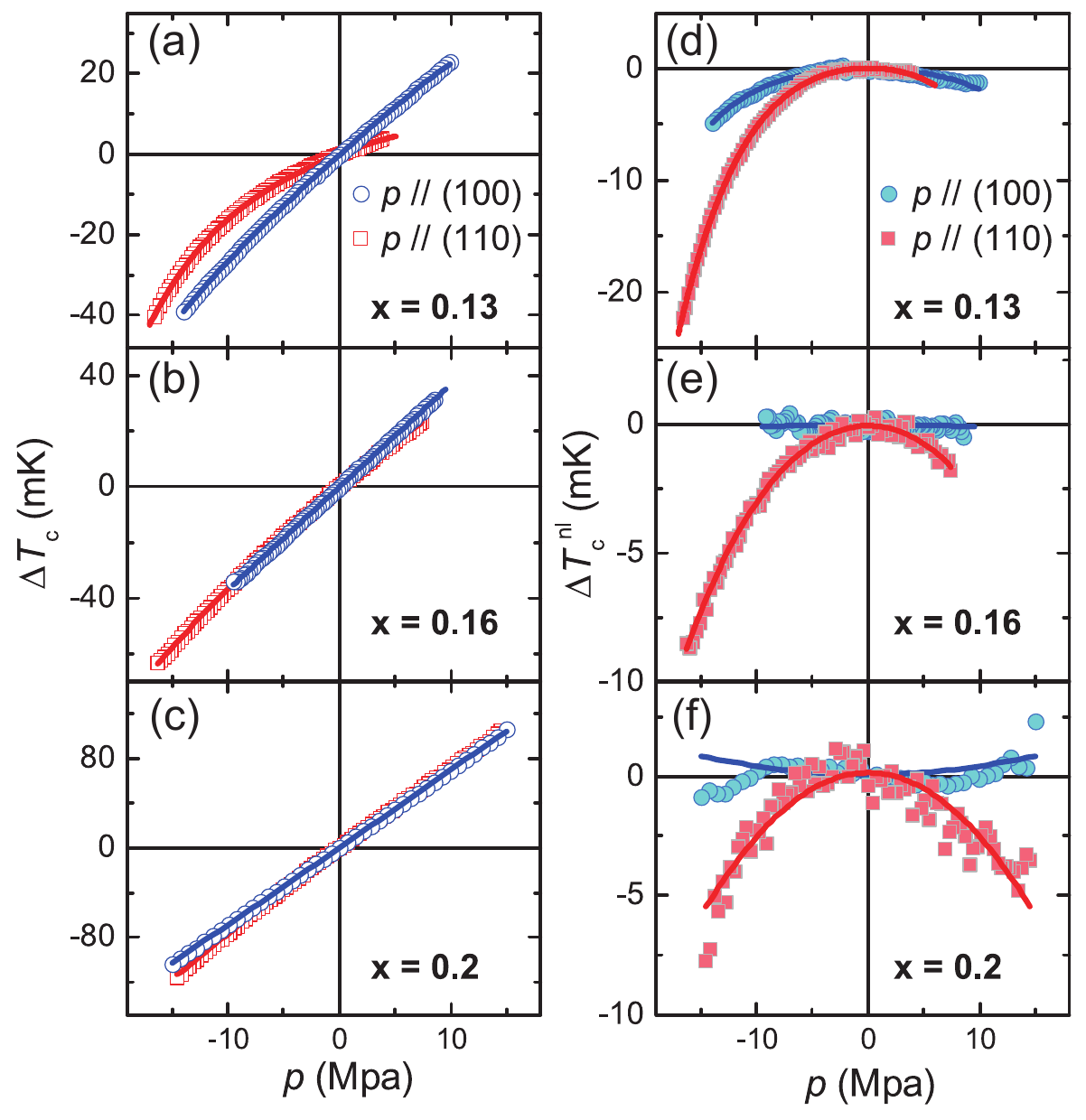}
\caption{(a)-(c) Uniaxial pressure dependence of $\Delta T_c$ along the (100) (circles) and (110) (squares) directions for $x$ = 0.13, 0.16, and 0.2, respectively. (d)-(f) Uniaxial pressure dependence of $\Delta T_c^{nl}$ for $x$ = 0.13, 0.16, and 0.2, respectively. The lines are the fitted results as described in the main text.}
\label{fig2}
\end{figure}

Single crystals of BaFe$_{2-x}$Ni$_x$As$_2$ (BFNA), Ba$_{0.67}$K$_{0.33}$Fe$_2$As$_2$ (BKFA), KFe$_2$As$_2$ (KFA) and LaFeAsO$_{0.74}$F$_{0.26}$ (LFAOF) were grown by the self-flux methods as reported previously \cite{ChenY11,WangM13}. The samples were cut into thin rectangular plates by a diamond saw along the desired directions determined by an x-ray Laue diffractometer. The tetragonal notation is used hereafter, i.e., the (110) and (100) directions correspond to the Fe-Fe and Fe-As-Fe directions, respectively. The uniaxial pressure dependence of the resistance was measured by a home-made uniaxial pressure device based on the piezo-bender as described previously \cite{LiuZ16,GuY17}. The piezobender of the uniaxial pressure device results in a slight hysteresis behavior between the processes of increasing and decreasing pressure due to its intrinsic properties, which is removed by averaging the pressures with the same resistance. The positive and negative values of pressure correspond to compressing and tensiling the samples, respectively.

\section{results and discussions}

The way of obtaining the pressure dependence of $T_c$ in this work is to measure the resistance change under the uniaxial pressure at various temperatures and then convert the data to the temperature dependence of resistance to calculate the $T_c$ at each pressure. Figures 1(a) and 1(b) show the results of the optimally doped BaFe$_{1.9}$Ni$_{0.1}$As$_2$ for uniaxial pressure along the (100) and (110) directions, respectively. For $p \parallel (100)$, the resistance $R$ is nearly linear with $p$ for the whole temperature range. Taking the values of $R$ at the same $p$, the temperature dependence of $R$ is shown in Fig. 1(c), which clearly demonstrates the change of $T_c$ under pressure. When the pressure is along the (110) direction, deviations from the linear behavior of $R(p)$ is observed around the superconducting transition. Again, the converted temperature dependence of $R$ is shown in Fig. 1(d).

Accordingly, the uniaxial pressure dependence $\Delta T_c$ can be derived as shown in Fig. 1(e), where $\Delta T_c$ = $T_c(p) - T_c(0)$, i.e., the relative change of $T_c$ to that at zero pressure. An almost linear relationship between the pressure and $T_c$ is observed for $p \parallel (100)$. When the pressure is along the (110) direction, $\Delta T_c$ shows clear nonlinear relationship with $T_c$. Here $T_c$ is determined as where $R$ becomes zero by the linear extrapolation of $R(T)$ during the transition. We have also tried to determine the value of $T_c$ by the onset and the middle temperature of the transition. The results are similar but with less certainty since the normal-state resistance along the (110) direction is affected by the uniaxial pressure \cite{LiuZ16}. 

\begin{figure}[t]
\includegraphics[width=\columnwidth]{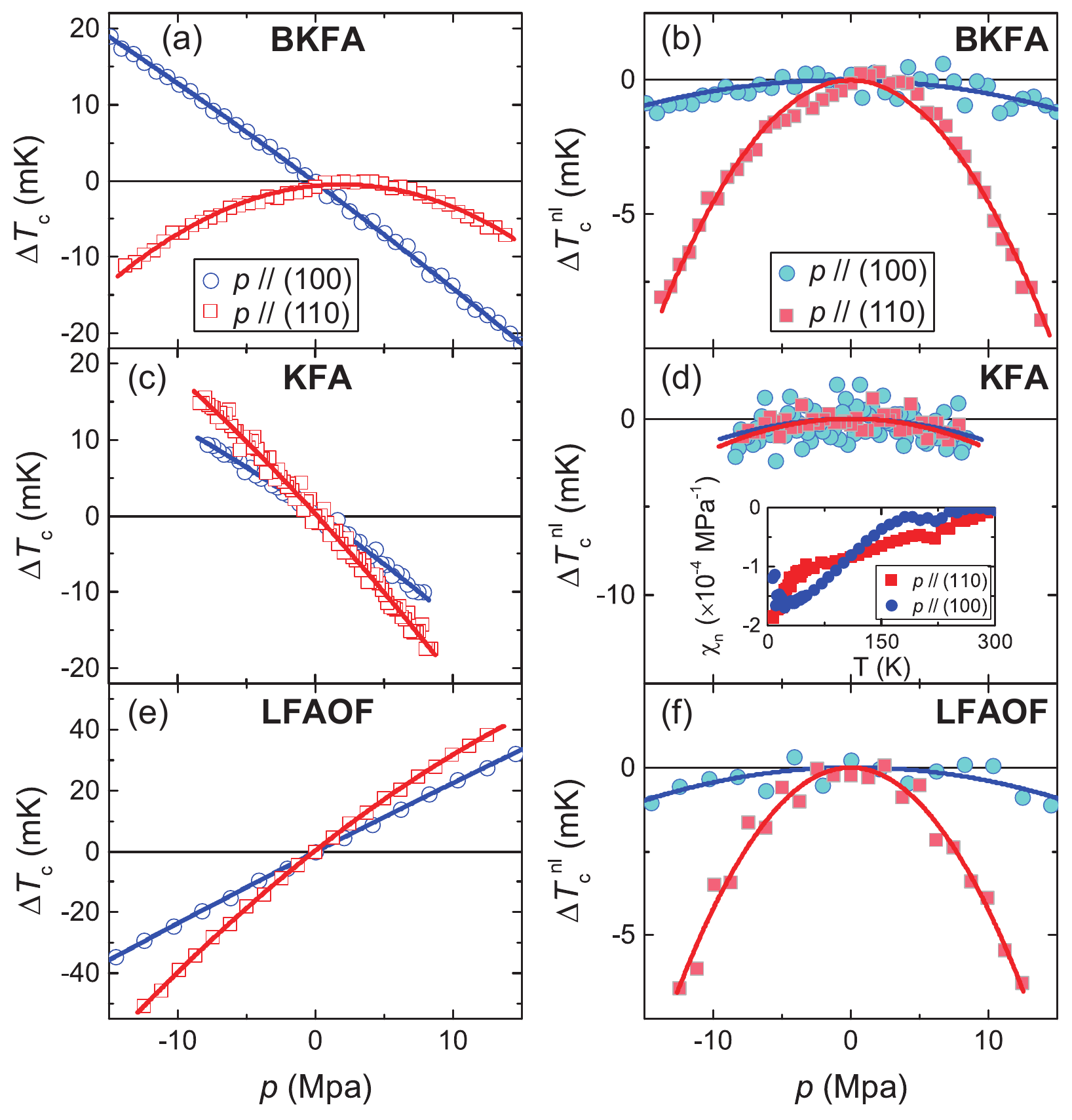}
\caption{ (a)-(f) Uniaxial pressure dependence of $\Delta T_c$ (left panels) and $\Delta T^{nl}_c$ (right panels) for the optimally doped Ba$_{0.67}$K$_{0.33}$Fe$_2$As$_2$  (a-b), KFe$_2$As$_2$ (c-d), and LaFeAsO$_{0.74}$F$_{0.26}$ (e-f). The circles and squares correspond to pressure along the (100) and (110) directions, respectively. The solid lines are fitted results as described in the main text. The inset in panel (d) shows the temperature dependence of $\chi_n$, which is proportional to the change of resistivity under $p$ as reported previously \cite{LiuZ16,GuY17}. }
\label{fig3}
\end{figure}

To quantitatively analyze the pressure dependence of $T_c$, we fit $\Delta T_c$ as $Bp + \Delta T^{nl}_c$, where $B$ is constant and $\Delta T^{nl}_c$ is the nonlinear component of $T_c(p)$. It is found that the following function is good enough to describe the data,
\begin{equation}
\Delta T^{nl}_c = Cp^2+Dp^4,
\label{Tc_eq}
\end{equation}
\noindent where $C$ and $D$ are all constants. Figure 1(f) shows the pressure dependence of $\Delta T^{nl}_c$, which clear shows the anisotropic behavior.

After having established the nonlinear pressure dependence of $T_c$ in optimally-doped BaFe$_{1.9}$Ni$_{0.1}$As$_2$, we further investigate overdoped compositions. 
Figures 2(a)-2(c) show $\Delta T_c$ for $x = 0.13$, $0.16$ and $0.2$ BaFe$_{2-x}$Ni$_x$As$_2$, respectively. With increasing doping, the contribution from the linear component of $T_c(p)$ increases significantly, which makes it hard to directly extract the nonlinear contribution. Yet, after subtracting the linear component, $\Delta T^{nl}_c$ still shows anisotropic behavior, as shown in Figs. 2(d)-2(f).

The nonlinear pressure dependence of $T_c$ is also observed in other iron-based superconductors. Figure 3(a) shows the results of optimally doped Ba$_{0.67}$K$_{0.33}$Fe$_2$As$_2$, whose anisotropic behavior of $\Delta T^{nl}_c$ is similar to that in BaFe$_{2-x}$Ni$_{x}$As$_2$; i.e., it is more significant along the (110) direction. For KFe$_2$As$_2$, the nonlinear behavior becomes weaker and isotropic, as shown in Fig. 3(c). The nonlinear behavior of $T_c$ in LaFeAsO$_{0.74}$F$_{0.26}$ is similar to that in Ba$_{0.67}$K$_{0.33}$Fe$_2$As$_2$, as shown in Fig. 3(e), although the isotropic contribution is larger.

\begin{figure}
\includegraphics[width=\columnwidth]{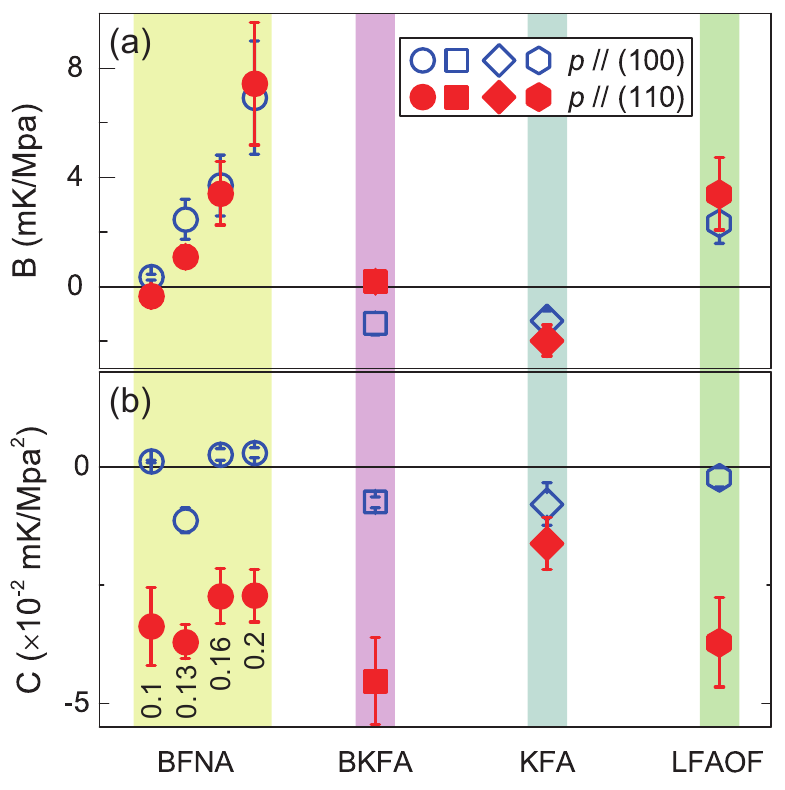}
\caption{The linear coefficient $B$ and the quadratic coefficient $C$ in BaFe$_{2-x}$Ni$_x$As$_2$ (circles), Ba$_{0.67}$K$_{0.33}$Fe$_2$As$_2$  (squares), KFe$_2$As$_2$ (diamonds) and LaFeAsO$_{0.74}$F$_{0.26}$ (hexagons). The open and solid symbols correspond to $p$ parallel to (100) and (110) directions, respectively. The error bars are estimated by considering the uncertainties from measurements of the cross sections, the zero-pressure positions and the fitting processes. }
\label{fig4}
\end{figure}

Figure 4(a) shows the fitting parameter $B$ in different samples. Since $B$ corresponds to the linear dependence of $T_c$ on the uniaxial pressure $p$, it is similar to $dT_c/dp$ in previous reports \cite{MeingastC12}. In BaFe$_{2-x}$Ni$_x$As$_2$, $B$ increases with increasing doping level, which is consistent with previous reports on Ba(Fe$_{1-x}$Co$_x$)$_2$As$_2$ \cite{MeingastC12}. Overall, the values of $B$ in BaFe$_{2-x}$Ni$_x$As$_2$, KFe$_2$As$_2$ and Ba$_{0.67}$K$_{0.33}$Fe$_2$As$_2$ are very close for pressure applied along the (110) and (100) directions, suggesting a nearly isotropic response of superconductivity to $p$. In optimally doped Ba$_{0.67}$K$_{0.33}$Fe$_2$As$_2$ , $B$ also shows a slightly anisotropic behavior, probably because the exact doping levels of the samples are slightly different due to the inhomogeneity in the growing process and $dT_c/dp$ changes sign around optimal doping \cite{BohmerAE15}. The values of the quadratic $C$ coefficient for different compounds are plotted in Fig. 4(b). Except for KFe$_2$As$_2$, the values of $C$ all show large anisotropic behavior, and are very similar for different samples when strain is applied along the (110) direction. Note that $C$ becomes isotropic and rather small in KFe$_2$As$_2$.

Our results demonstrate that the response of superconductivity to the uniaxial pressure within the $ab$ plane is composed of linear and nonlinear components. The latter can be described as an even function of the pressure, where the quadratic term dominates. Moreover, the quadratic coefficient along the (110) direction is usually larger than that along the (100) direction. The nonlinear behavior seems to be unique for iron-based superconductors since it is not found in a cuprate sample, as shown in the appendix. As the (110) direction is associated with the nematic order direction and the uniaxial pressure acts as an external field to the nematic order \cite{LiuZ16}, we propose that the nonlinear response of $T_c$ to the uniaxial pressure is due to the coupling between superconductivity and nematic fluctuations. Indeed, the change of resistivity under $p$ along the (110) direction is usually much larger than that along the (100) direction \cite{HosoiS16,LiuZ16}. This also explains why the quadratic coefficient $C$ is small and nearly isotropic in the case of KFe$_2$As$_2$, since this compound seems to be far from a nematic instability and, as shown by the inset of Fig. 3(d), its resistance response to uniaxial pressure is nearly isotropic already in the normal state.

A phenomenological symmetry analysis sheds important light on the behaviors observed here \cite{FernandesRM13a,FernandesRM13b}. Pressure along the (110) direction induces not only shear strain in the $B_{2g}$ channel, $\varepsilon_{B_{2g}} =  \partial_x u_y + \partial_y u_x$, where $\vec{u}$ is the displacement vector, but also strain in the other symmetry channels due to the finite Poisson ratio  \cite{Fisher17} -- including the isotropic strain $\varepsilon_{A_{1g}} = \partial_x u_x + \partial_y u_y$. The latter couples to the square of the superconducting order parameter $\Delta$ in a Landau free energy expansion, resulting in the linear dependence of $T_c$ on $p$. On the other hand, $\varepsilon_{B_{2g}}$ acts as a conjugate field to the nematic order parameter $\varphi$, inducing a finite value $\varphi \propto \varepsilon_{B_{2g}} \chi_n$ that can be sizable if the nematic susceptibility $\chi_n$ is large -- as expected near a (quantum) nematic phase transition. Now, because $\Delta$ and $\varphi$ have a biquadratic coupling in the Landau free energy, $T_c$ acquires a quadratic dependence on $p$. The fact that the quadratic coefficient $C$ in Eq. (\ref{Tc_eq}) is negative implies that this biquadratic coefficient is positive, i.e., nematicity and superconductivity compete with each other. The additional quartic coefficient $D$ in Eq. (\ref{Tc_eq}) is likely a consequence of the relatively large pressures applied experimentally. 

On the other hand, pressure along the (100) direction induces both $\varepsilon_{A_{1g}} = \partial_x u_x + \partial_y u_y$ and $\varepsilon_{B_{1g}} =  \partial_x u_x - \partial_y u_y$. The fact that the linear coefficient $B$ in $T_c(p)$ is essentially the same for both uniaxial pressure directions implies that the induced $\varepsilon_{A_{1g}}$ strain is nearly the same in both cases. In contrast, the very small quadratic coefficient $C$ in the case of $p \parallel (100)$ can be attributed to the absence of nematicity in the $B_{1g}$ channel, i.e., the $B_{1g}$ nematic order parameter induced by $\varepsilon_{B_{1g}}$ is small.

Our results suggest that the interplay between nematic fluctuations and superconductivity is ubiquitous in iron-based superconductors, which is consistent with previous results that nematic fluctuations are present above $T_c$ in various systems \cite{KuoHH16,HosoiS16,ThorsmolleVK16,LiuZ16,GuY17}. Thus, elucidating the superconducting state in these systems likely requires understanding the effects of nematic fluctuations. Within the framework of the above analysis, the negative values of $C$ in Fig. 4 reveal the competition between the nematic and superconducting order parameters. It is surprising that $C$ changes little with doping in BaFe$_{2-x}$Ni$_x$As$_2$ and is still observable in KFe$_2$As$_2$, since one would expect the nematic fluctuations to be weak in these samples as they are far away from optimally doping levels. It should be pointed out that we have no reliable way to obtain the amplitude of nematic fluctuations in overdoped samples, but previous works indicate that it increases with increasing doping in the underdoped regime \cite{GuY17}. Whether the doping-dependence of $C$ is compatible with the existence of nematic quantum critical fluctuations, which have been shown to exist in many systems \cite{KuoHH16,HosoiS16,ThorsmolleVK16,LiuZ16,GuY17} and thought to be important to superconductivity \cite{Yamase13,LedererS15,MetlitskiMA15,SchattnerY16,PhilippTD16,LedererS17,Kang16}, remains to be established. Importantly, the interplay between nematic fluctuations and superconductivity is by no means limited to the form we have discussed here. Future experiments and theories are desired to further elucidate these issues.

\section{conclusions}

In conclusion, our results provide direct evidence for the coupling between nematicity and superconductivity by revealing the in-plane anisotropic behavior of $T_c(p)$. A quadratic $p$ dependence of $T_c$ is found and can be explained by the biquadratic coupling between the superconducting and nematic order parameters. The fact that it is ubiquitous in iron-based superconductors indicates the importance of nematic fluctuations for superconductivity. 
 
Experimental work is supported by the National Key R\&D Program of China (No. 2017YFA0302903, No. 2016YFA0300502,  No. 2018YFA0704201, No. 2017YFA0303103, No. 2015CB921302, No. 2015CB921303), the National Natural Science Foundation of China (No. 11674406, No. 11874401, No. 11961160699, No. 11374346, No. 11374011, No. 11774401, No. 11522435, and No. 11822411), the “Strategic Priority Research Program(B)” of the Chinese Academy of Sciences (No. XDB25000000, No. XDB07020000 and No. XDB28000000), and China Academy of Engineering Physics (No. 2015AB03). H. L. and Y. Y. are supported by the Youth Innovation Promotion Association of CAS. Theoretical work by R.M.F was supported by the U.S. Department of Energy, Office of Science, Basic Energy Sciences, under Award No. DE-SC0012336.

\appendix
\section{The coefficient $D$}

\begin{figure}
\includegraphics[width=0.8\columnwidth]{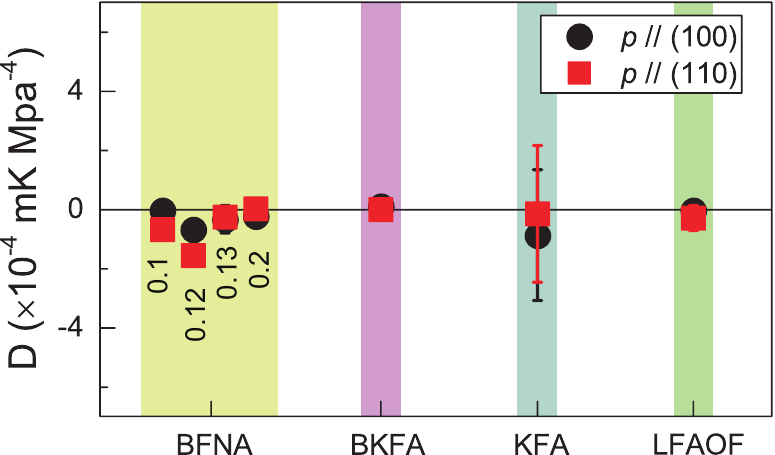}
\caption{The quartic coefficient $D$ in BFNA, BKFA, KFA and LFAOF. The circles and square symbols correspond to uniaxial pressure $p$ parallel to (100) and (110) directions, respectively.}
\label{figS1}
\end{figure}

The coefficients $D$ in Eq. (1) in main text for different iron-based superconductors are shown in Fig. \ref{figS1}. The abbreviations for the samples are the same as in the main text. The values of $D$ are all near zero and change little with different doping levels and systems. This suggests that the quartic term has smaller effect on $\Delta T_c^{nl}$ than quadratic term for small pressures. 

\section{Measurements on cuprate Bi-2212}

We performed the same resistance measurements on single crystals of optimally doped copper-oxide high-temperature superconductors Bi$_{2-x}$Pb$_x$Sr$_2$CaCu$_2$O$_{8+\delta}$ (Bi-2212). High-quality single crystals of Bi-2212 were grown by the traveling solvent floating zone technique. Note that, for Bi-2212, the orthorhombic notation is used in the experiments. The (100) and (110) directions are for Cu-O-Cu and diagonal directions, respectively. The electronic contacts were made by standard a two-part silver paste with heating up at 350$^{\circ}$ for 2 h. The typical contact resistance is less than 5 $\Omega$. Other sample preparation and measuring methods were similar as those in iron-based superconductors depicted in main text. 

\begin{figure}
\includegraphics[width=\columnwidth]{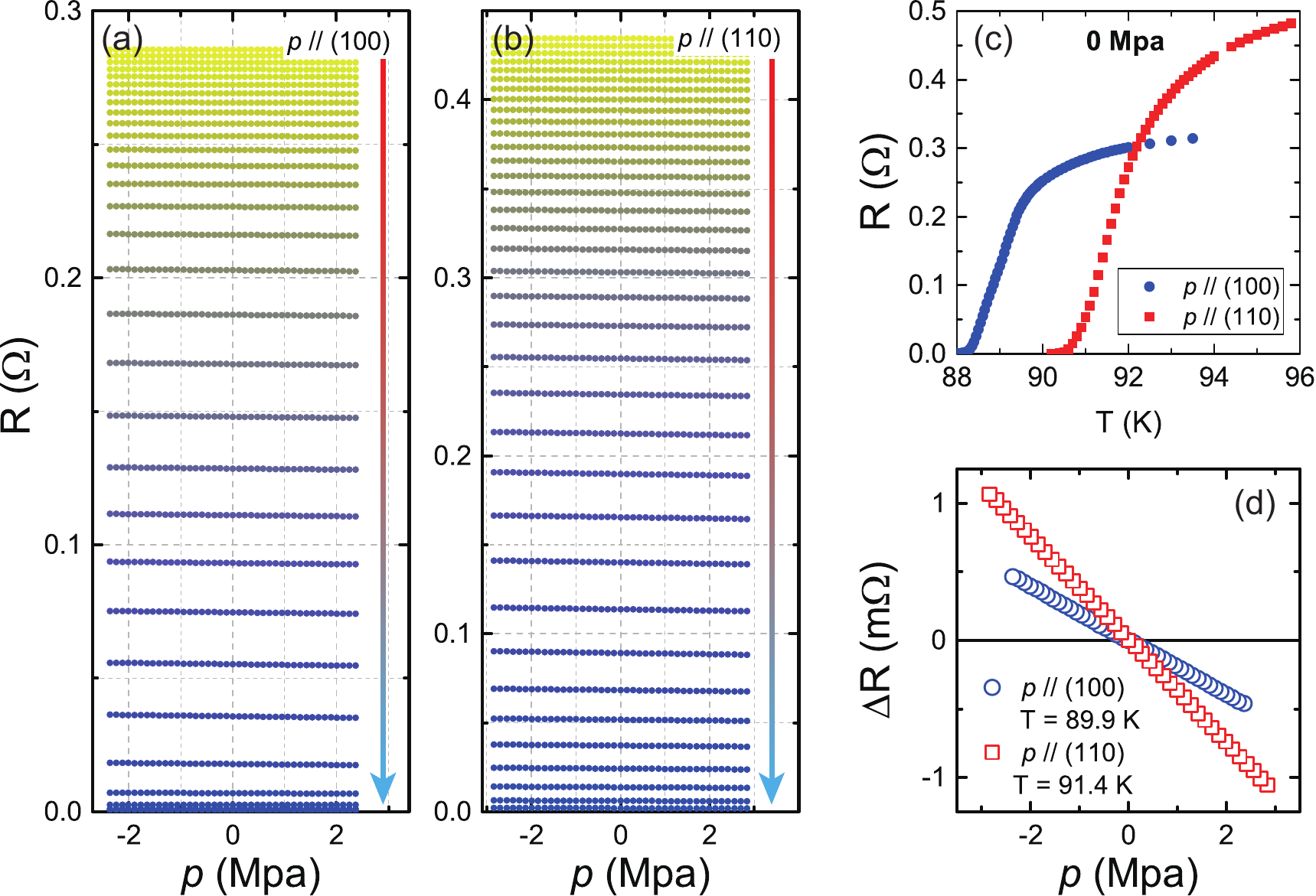}
\caption{(a) Uniaxial pressure dependence of resistance for optimal doping Bi-2212 with pressure $p$ along the (100) direction near superconducting transition temperature. The temperature varies from 91 to 88 K with a step of 0.1 K indicated by right gradient arrow. (b) Similar to panel (a) but for $p$ along (110) direction. Temperature varies from 94 to 90.5 K with 0.1 K step. Note that the orthorhombic notation is used. (c) Temperature dependence of resistance subtracted from panels (a) and (b) at $p=0$ Mpa. Although the two samples are cut from same batch, the $T_c$'s are slightly different which mainly comes from their different current densities. (d) Uniaxial pressure dependence of $\Delta R$ with pressure along (100) (blue circles) and (110) (red squares) directions. The specific temperatures are the maximum temperature of $dR/dT$ with $T=$ 89.9 and 91.4 K, respectively.}
\label{figS2}
\end{figure}

Here, since the flake of Bi-2212 single crystals are very fragile under uniaxial pressure, we measured them with quite small pressure range, i.e., less than about $\pm$3 Mpa. As shown in Fig. \ref{figS2}(a) and \ref{figS2}(b), the uniaxial pressure dependence of resistance in either (100) or (110) directions presents no obvious curvature with temperature range from normal state to superconducting state, which is quite different from the nonlinear behaviors in other iron-based superconductors near optimal dopings. To clearly show the linear dependence on uniaxial pressure of resistance, we display $\Delta R$ at middle transition temperatures for each samples in Fig. \ref{figS2}(d). The statistics R$^2$'s of linear fitting for both data exceed 0.9999. The temperature dependence of resistance at 0 Mpa for both samples (Fig. \ref{figS2}(c)) are converted from data in Figs. \ref{figS2}(a) and \ref{figS2}(b). The sharp superconducting transitions demonstrate high homogeneity in our cuprate samples. Although the two samples measured along different directions were cut from same crystal rod, the $T_c$'s are slightly different ($\sim$2 K) which were mainly caused by different current densities during the measurements due to different sample cross sections. 

\begin{figure}
\includegraphics[width=\columnwidth]{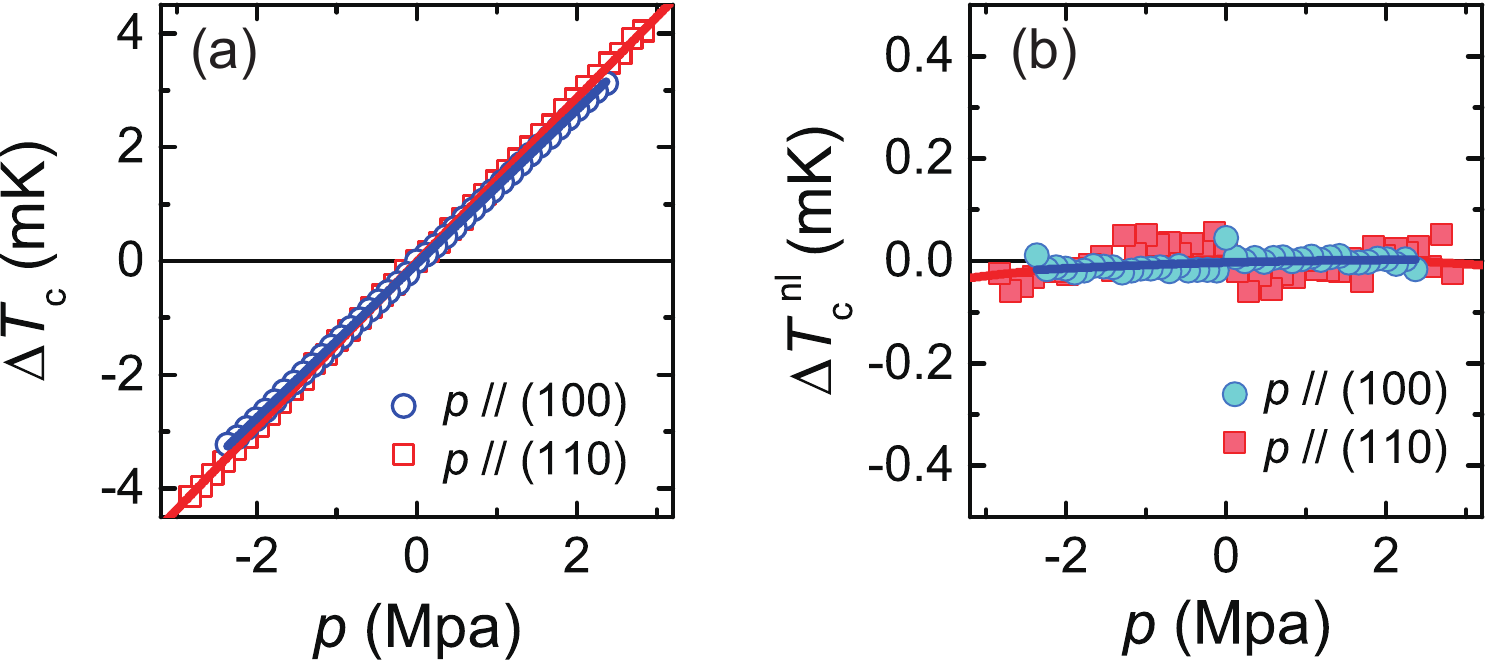}
\caption{(a) Uniaxial pressure dependence of $\Delta T_c$ along (100) (circles) and (110) (squares) directions. (b) Uniaxial pressure dependence of $\Delta T_c^{nl}$ along (100) (circles) and (110) (squares) directions.}
\label{figS3}
\end{figure}

We obtained the uniaxial pressure dependence of $\Delta T_c$ and $\Delta T_c^{nl}$ in Figs. \ref{figS3}(a) and \ref{figS3}(b), respectively. Both $\Delta T_c$ show perfect linear relationship with the uniaxial pressure and $\Delta T_c^{nl}$ is extremely small compared with results in other iron-based superconductors in the main text. Therefore, there is no nonlinear behavior of $T_c$ under uniaxial pressure in Bi-2212, at least for small pressure.

\end{document}